\documentclass[aps,twocolumn,floatfix,superscriptaddress]{revtex4}

\usepackage{graphicx}
\usepackage{lscape}
\usepackage{indentfirst}
\usepackage{latexsym}
\usepackage{multirow}
\usepackage{tabls}
\usepackage{epsfig}
\usepackage{multirow}
\usepackage{subfigure}
\usepackage{hyperref}

\usepackage{color}
\usepackage{amssymb}

\usepackage{amsfonts}
\usepackage{amsmath}
\usepackage{bm}
\usepackage{mathrsfs}

\newcommand{\ratio}{m_1/m_2}

\def\lsim{\mathrel{\rlap{\lower3pt\hbox{\hskip1pt$\sim$}}
    \raise1pt\hbox{$<$}}}                
\def\gsim{\mathrel{\rlap{\lower3pt\hbox{\hskip1pt$\sim$}}
    \raise1pt\hbox{$>$}}}         

\def\coordeq{ \, \mathrel{ \rlap{\hbox{\hskip-2.5pt$=$} }
    \raise4pt\hbox{$\cdot$}} \, }                

\begin{document}

\title{Towards beating the curse of dimensionality for gravitational waves using Reduced Basis}

\def\addBrownPhys{Department of Physics, Brown University, Providence, RI 02912, USA}
\def\addJPL{Jet Propulsion Laboratory, California Institute of Technology, Pasadena, CA 91109 USA}
\def\addCaltech{TAPIR, California Institute of Technology, Pasadena, CA 91125, USA}
\def\addUMD{Department of Physics, Maryland Center for Fundamental Physics, Joint Space Sciences Institute, Center for Scientific Computation and Mathematical Modeling, University of Maryland, College Park, MD 20742, USA}
\def\addBrownMath{Division of Applied Mathematics, Brown University, Providence, RI 02912, USA}
\def\addUWM{Center for Gravitation and Cosmology, University of Wisconsin-Milwaukee, Milwaukee, WI 53201, USA}
\def\addUMDEO{Department of Physics, Maryland Center for Fundamental Physics, University of Maryland, College Park, MD 20742, USA}

\author{Frank Herrmann}
\affiliation{\addUMD}

\author{Scott E. Field}
\affiliation{\addUMD}

\author{Chad R.\,Galley}
\affiliation{\addJPL}
\affiliation{\addCaltech} 

\author{Evan Ochsner}
\affiliation{\addUWM}

\author{Manuel Tiglio}
\affiliation{\addUMD}

\begin{abstract}
	Using the Reduced Basis approach, we efficiently compress and accurately represent the space of waveforms for non-precessing binary black hole inspirals, which constitutes a four dimensional parameter space (two masses, two spin magnitudes). Compared to the non-spinning case, we find that only a {\it marginal} increase in the (already relatively small) number of reduced basis elements is required to represent any non-precessing waveform to nearly numerical round-off precision. Most parameters selected by the algorithm are near the boundary of the parameter space, leaving the bulk of its volume sparse. Our results suggest that the full eight dimensional space (two masses, two spin magnitudes, four spin orientation angles on the unit sphere) may be highly compressible and represented with very high accuracy by a remarkably small number of waveforms, thus providing some hope that the number of numerical relativity simulations of binary black hole coalescences needed to represent the entire space of configurations is not intractable.
Finally, we find that the {\it distribution} of selected parameters is robust to different choices of seed values starting the algorithm, a property which should be useful for indicating parameters for numerical relativity simulations of binary black holes. In particular, we find that the mass ratios $m_1/m_2$ of non-spinning binaries selected by the algorithm are mostly in the interval $[1,3]$ and that the median of the distribution follows a power-law behavior $\sim (m_1/m_2)^{-5.25}$.
\end{abstract}

\maketitle

\section{Introduction}

The upcoming generation of advanced-sensitivity ground-based gravitational wave interferometer detectors (i.e., advanced LIGO, advanced Virgo, Indigo, and KAGRA)
~\cite{LIGO_web,VIRGO_web,GEO_web,KAGRA_web} brings an increasing demand to accurately and efficiently represent gravitational waveforms from generic precessing compact binary sources~\cite{Abadie:2010cfa,Apostolatos_etal:PRD49,Brown:2012gs,Ajith:2011ec,Centrella:2010zf,Buonanno_etal:PRD70,Buonanno:2002fy,Pan_etal:PRD69}. Such waveforms for quasi-circular inspirals are parameterized by a set of eight intrinsic physical quantities -- two masses, two spin magnitudes, and four spin orientation angles on the unit sphere (8D) \footnote{Alternative theories of gravity may have more parameters than General Relativity; see, for example, \cite{Chatziioannou:2012rf,Yunes:2009ke} and references therein.}. 

The large dimensionality of this parameter space makes gravitational wave searches, parameter estimation, and modeling prohibitively expensive and computationally unfeasible with most methods. This problem is called the ``curse of dimensionality'' \cite{Bellman:2003:DP:862270} and, in particular, is a major hurdle for modeling astrophysical gravitational wave sources. In this setting, the cost of numerical relativity simulations of the full Einstein equations describing the inspiral, merger, and ringdown of binary black hole coalescences is so expensive that an optimal or nearly optimal criterion for selecting which points in parameter space to simulate is thus critical. This is also the case when numerical simulations are used to calibrate or build effective one body \cite{PhysRevD.79.081503,Buonanno:1998gg,Pan:2009wj,Taracchini:2012ig} or phenomenological \cite{Ajith:2007qp,Ajith:2009bn,Santamaria:2010yb}  models. 

To address these issues, the Reduced Basis (RB) approach was introduced to gravitational wave physics in  Refs.~\cite{Field:2011mf,Caudill:2011kv} and shown to efficiently compress the space of waveforms with a very small loss of accuracy, typically of the order of numerical round-off, for any given bounded parameter domain. The compression is accomplished by determining a set of nearly optimal physical parameter values from which a basis is constructed to represent any given waveform within the same physical model  through its projection onto this basis. Details of the algorithm, which is known as the {\it greedy algorithm} can be found in \cite{Field:2011mf, Caudill:2011kv}.

The Reduced Basis approach has several uniquely appealing features such as its hierarchical compression of the waveform space, its ability to handle large numbers of parameters, and its ability to identify the most relevant points in the parameter space. The latter may be particularly important for guiding which numerical relativity simulations to perform in order to calibrate effective-one-body or to fit phenomenological models. 

In this paper we apply the RB approach to {\it non-precessing} binary black hole inspirals that are non-spinning (2D), have equal spin magnitudes (3D), or have unequal spin magnitudes (4D). To our knowledge, this is the first paper, along with Refs.~\cite{Caudill:2011kv,Cannon:poster}, applying reduced order modeling techniques to gravitational waves with a larger number of parameters than previously considered. 

The results presented in this paper point towards the possibility  that the eight-parameter space of waveforms for precessing binary inspirals might admit a remarkably compact representation. For the non-precessing inspiral waveforms, we find a very moderate increase in the number of RB elements as the dimensionality of the parameter space is increased from 2D to 4D. While this number could and might change significantly for precessing binaries, former results in \cite{Galley:2010rc} show dimensionality reduction in the precessing case (with respect to the spin dynamics).

\section{Spin-Aligned Post-Newtonian Waveforms}\label{sec:pneq}
Throughout this paper we use the so-called restricted TaylorF2 post-Newtonian (PN) waveforms (see~\cite{Buonanno:2009zt} and references therein for a more detailed discussion of TaylorF2 and other PN approximants). 
These waveforms use the stationary-phase approximation to construct analytic frequency-domain waveforms of the form
\begin{equation}
\tilde{h}(f) = {\cal A} f^{-7/6} e^{i\,\Psi(f)}\, , 
\end{equation}
where $\Psi(f)$ is expressed as a polynomial in the PN expansion parameter $v = (\pi M f)^{1/3}$ and ${\cal A}$ is a constant such that the $\tilde{h}$ is normalized to unity.
Spin-independent corrections to $\Psi(f)$ are currently known to $v^7$, or 3.5PN order~\cite{Blanchet:2001ax}. The dimensionless spin vectors of the component masses, $\vec{\chi}_i = \vec{S}_i / m_i^2$ with $0 \leq |\vec{\chi}_i| \leq 1$, can also enter the corrections to $\Psi(f)$ and these spin-dependent corrections are known to 2.5PN order. 
These include spin-orbit corrections of the form $\vec{\chi}_i \cdot \hat{L}_N$ (where $\hat{L}_N$ is the ``Newtonian'' orbital angular momentum, which is transverse to the orbital plane) at 1.5PN and 2.5PN order, and spin-spin and self-spin corrections of the forms $\vec{\chi}_1 \cdot \vec{\chi}_2$ and $\vec{\chi}_i^{\,2}$ at 2PN order ~\cite{Kidder:1995zr,Faye:2006gx,2006PhRvD..74j4034B, Arun:2008kb}.

For a generic binary system the spins and orbital plane precess, so the relative orientations among $\vec{\chi}_1$, $\vec{\chi}_2$ and $\hat{L}_N$, and thus the spin-dependent PN corrections of $\Psi(f)$, vary on a precessional timescale. 
Currently, there are no frequency-domain template families for precessing spins, although there is at least one such effort underway~\cite{AjithFavata}. In the ``spin-aligned'' case, when each spin is aligned or anti-aligned with $\hat{L}_N$, the binary does not precess and the relative orientations of $\vec{\chi}_1$, $\vec{\chi}_2$ and $\hat{L}_N$ (and the spin-dependent corrections to $\Psi(f)$) remain constant throughout inspiral. Thus the simple TaylorF2 model is valid and does not need to be augmented with precession equations. The spin-dependent corrections are expressed in terms of the (constant) projection of each spin along the orbital angular momentum: $-1 \leq \chi_i \equiv \vec{\chi}_i \cdot \hat{L}_N \leq 1$.

In the \emph{restricted} TaylorF2 approximation, the amplitude is expanded only to leading order while the phase is expanded to a higher order. In all cases, the spin-independent contributions to the phase are here included up to 3.5PN order, as given by Eq. (3.18) of~\cite{Buonanno:2009zt}. In cases where spin is included, the 3.5PN non-spinning phase is augmented with spin-orbit, spin-spin and self-spin corrections through 2.5PN order, as can be read from Eqs. (6.22)-(6.25) of~\cite{Arun:2008kb}.

\section{Results}

We consider binary black holes \footnote{Because binary neutron stars can have lower component masses, possibly in the range $[1,3] M_\odot$, there are many more cycles in band in the corresponding waveforms when compared to those of binary astrophysical black holes. This introduces a larger offline computational cost to build  Reduced Bases, but not a conceptual one. Hence, we use binary black holes for the foundation of our studies here and defer a detailed study of binary neutron stars once some additional computational developments to alleviate such offline costs are in place (see the discussion at the end of Section~\ref{sec:remarks}).} with individual component masses $m_i \in [3, 30] M_\odot$ and dimensionless spin magnitudes $\chi_i \in [-1, 1]$. Each mass direction (or dimension) in the training space (any discretization of the continuum space of parameters) is sampled with $n_m$ points and each dimensionless spin magnitude direction with $n_s$ points. After some numerical experimentation, we found that the majority of the selected mass components have values at the lower end of the considered range. Thus, we use a training space with points logarithmically spaced in the $m_1$-$m_2$ plane to provide a sufficiently dense cluster of points in the training space at those lower mass values. The inner products between any two waveforms are 
weighted by the reciprocal of the power spectral density (PSD) for advanced LIGO \footnote{As briefly discussed in \cite{Field:2011mf}, the form of the PSD is largely irrelevant for our purposes since our ability to represent waveforms with very high accuracy allows for one to represent waveforms of a template bank for other detectors, usually with very little loss of accuracy.} as given by the fitting formula in Ref.~\cite{Ajith:2009fz}, with a lower cutoff frequency of $10$Hz and a maximum one corresponding to the highest ISCO frequency. For clarity, when we refer to the number of RB elements we mean the number of basis vectors needed to represent the training space to within the specified tolerance, typically round-off ($\sim 10^{-14}-10^{-12}$). The number of RB elements also equals the number of selected parameter tuples of mass and spin magnitudes.

Compared to the non-spinning case, the extra dimensionality of the spin parameter space (for non-precessing inspiral waveforms, as considered here) has a remarkably small effect on the final number of RB elements needed to  represent the {\it entire} training space with numerical round-off precision. In fact, as will be discussed in Section \ref{sec:reconstruct}, the same RB represents {\it any} waveform, and not necessarily a member of the training space, in the same range of parameters to within essentially round-off precision. These are the most important results of this paper.

\subsection{Effect of increasing dimensionality}\label{sec:dimens}

\begin{figure}[htp]
  \includegraphics[width=\linewidth]{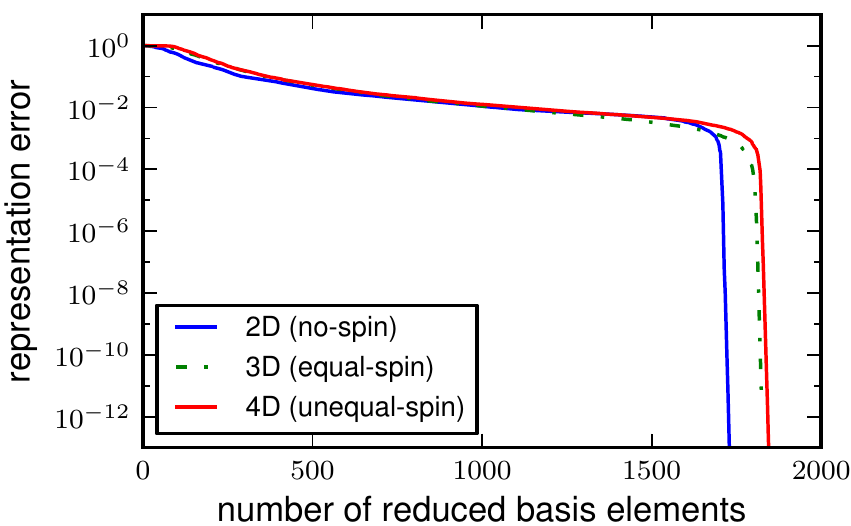}
  \caption{The reduced basis representation error as a function of the number of reduced basis for different dimensionalities. The greedy error shows the rapid exponential convergence to round-off. The number of reduced basis elements increases very mildly from the 2D (no-spin) to the 3D (equal-spin) case and to the 4D (generic non-precessing spins) case. This shows that the increasing dimensionality may not be a major obstacle for the Reduced Basis approach.}
  \label{fig:dimensionality}
\end{figure}

Figure~\ref{fig:dimensionality} shows the maximum representation error for training spaces, corresponding to 2D, 3D, and 4D parameter spaces of non-precessing inspiral waveforms, as a function of the number of RB elements. For each waveform $\tilde{h}$ in the training space we compute the representation error\footnote{$P_{N}[\tilde{h}]$ is the best representation of $\tilde{h}$ using $N$ reduced basis elements, see Ref.~\cite{Caudill:2011kv} for more details.} $\|\tilde{h}-P_{N}[\tilde{h}] \|^2$ and then plot its maximum over the entire training space, as the number of RB elements is increased.
In all cases Fig.~\ref{fig:dimensionality} shows a similar behavior, namely, an initially slow fall-off in the representation error and a later rapid convergence to round-off. 

The 2D  runs were done with up to $n_m=400$ points in each mass dimension, for a maximum size of the training space of $n_m^2=1.6 \times 10^5$ total samples. For the 3D (4D) runs we used up to $n_m=200$ ($100$)  and  $n_s=50$ ($20$) samples in the spin direction for a training space with a maximum size of $2 \times 10^6$ ($4 \times 10^6$) samples. 
In order to directly compare the 3D and 4D cases in Fig.~\ref{fig:dimensionality}, we use $n_m=100$ samples for both since, as discussed in Section~\ref{sec:reconstruct}, more samples are not needed for high accuracy reconstruction of arbitrary 3D or 4D waveforms.

The key result from Fig.~\ref{fig:dimensionality} is the very moderate increase in the maximum number of RB elements needed as the dimensionality increases from 2D to 3D to 4D. At an error of $10^{-11}$ the 2D case requires $\approx 1,\!725$, the 3D case $\approx 1,\!824$, and the 4D case $\approx 1,\!839$ basis vectors. The number of basis vectors that result by increasing the dimensionality of the parameter space from 2D to 4D is only a $6.6\%$ increase and from 3D to 4D requires only $15$ more basis vectors. Despite doubling the dimensionality of the parameter space, the number of basis vectors needed to represent the four-parameter waveform space is nearly the same as for non-spinning binaries. 

Naive scaling arguments would suggest an increase in the number of RB elements by orders of magnitude. If the 2D case requires $\sim 2 \times 10^3$ basis vectors then one might be lead to think that doubling the number of parameters could require more by a factor the size of the training space in the spin directions, which would here be about $20^2$ for a total of $\sim 8 \times 10^5$. The fact that we observe an exceedingly small relative increase in the number of RB elements lends hope that the number of basis vectors needed to represent the full 8D parameter space for inspiral waveforms is also much smaller than what might be anticipated by estimating the volume of the parameter space or by using naive (such as equally spaced or random) sampling techniques. This is only suggestive since precession has significant effects on the structure of the inspiral waveforms~\cite{Apostolatos_etal:PRD49,Kidder:1995zr,Arun:2008kb}. But if it holds, the curse of dimensionality might be beaten. In fact, indications that the problem, in the presence of precession, is amenable to dimensional reduction, have already been found through a Principal Component Analysis of the precessing dynamics of compact binary inspirals \cite{Galley:2010rc}. There it was found that for the case of a random selection of precessing binaries with the same total mass there are three combinations of spin orientations that are semi-conserved (in a statistical sense) throughout the inspiral. The presence of such combinations implies that the dynamical configurations can be parameterized by a reduced number of independent parameters than the fiducial seven (1 mass, 6 spin components).

\subsection{Parameter values selected}

The Reduced Basis-Greedy Algorithm approach possesses several unique features. One of them is being able to identify, in a precise mathematical sense
, a nearly optimal set of points in the physical parameter space \cite{Binev10convergencerates}. In this section we analyze  the structure of the selected parameter values for the models and scenarios here studied.

\subsubsection{Mass parameters (2D, 3D, 4D)}\label{sec:masspars}

\begin{figure}[htp]
  \includegraphics[width=\linewidth]{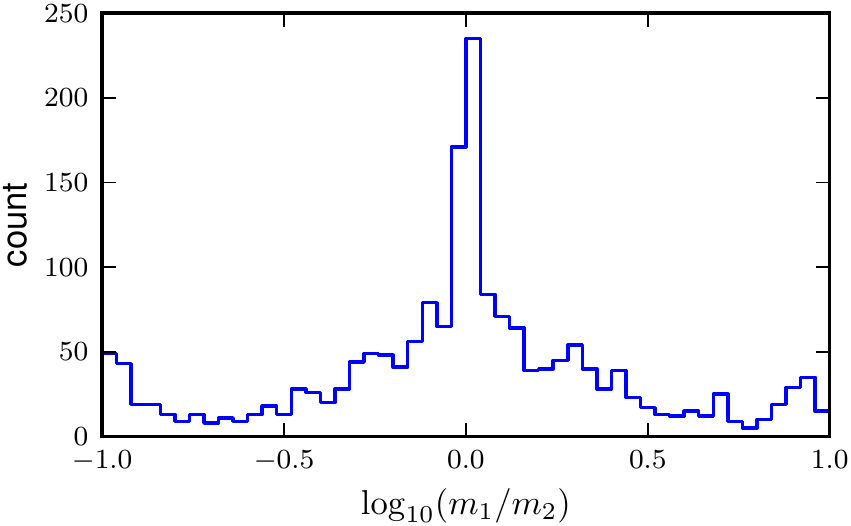}
  \caption{Histogram of the selected parameter choices for the mass-ratio $\ratio$ in the 4D case. Note the large spike near equal masses and the rapid fall-off to small counts for increasing disparity in the component masses.}
  \label{fig:qhist}
\end{figure}

\begin{figure}[htp]
  \includegraphics[width=\linewidth]{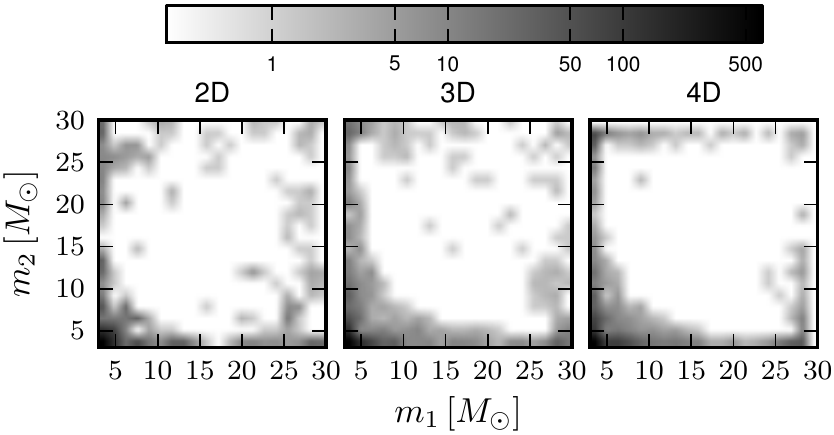}
  \caption{The reduced basis parameter choices in the $m_1$-$m_2$ plane. The overall structure is similar for all three (2D, 3D, 4D) non-precessing cases. The grey-scale bar utilizes a logarithmic scaling.}
  \label{fig:m1m2par}
\end{figure}

Figure~\ref{fig:qhist} shows the distribution of selected points for the mass ratio $\ratio$ in the non-precessing (4D) case. There we plot $\log_{10}(\ratio)$ to obtain a symmetric representation in $m_1$ and $m_2$. Since in this paper we focus on the range $m_1, m_2\in [3,30] M_{\odot}$, the mass-ratio has a range $\ratio\in[1/10,10]$. The selected values by the greedy algorithm have a strong peak at the equal-mass case. This follows from the selected individual mass components, which are clustered at low values (see also Fig.~\ref{fig:m1m2par}).

We have found similar results for the distribution of $\ratio$ in the 2D and 3D cases, also resulting from the low-mass selection. This is also visible from Fig.~\ref{fig:m1m2par}, which shows the selected parameter values in the $m_1$-$m_2$ plane.   
Notice that there is little qualitative difference in going from 2D to 3D to 4D in Fig.~\ref{fig:m1m2par}. This indicates that the problem might admit some kind of dimensional separability. For example, it might be possible to initially select $(m_1,m_2)$ values through a 2D greedy strategy and then use only those to populate the $m_1$-$m_2$ plane in the construction of very efficient and compact training spaces for building reduced basis in higher dimensional parameter cases. This is beyond the scope of this paper, but it might provide a dramatic reduction in training space sizes and, if so, a possible avenue for expansions to higher dimensions and the full 8D problem.

\subsubsection{Mass and spin parameters (3D)}

\begin{figure}[htp]
  \includegraphics[width=\linewidth]{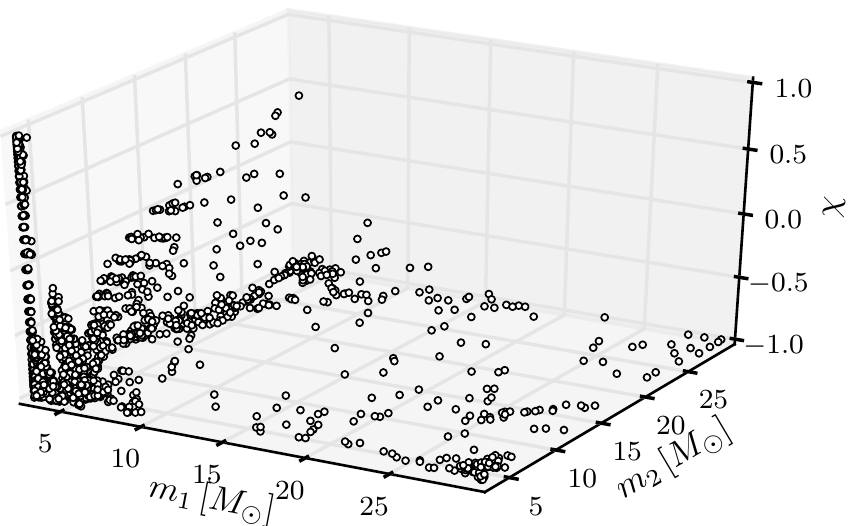}
  \caption{The reduced basis parameter choices for the 3D case $(m_1, m_2, \chi)$. Comparing to Fig.~\ref{fig:m1m2par} one can also see the selection of primarily low-mass systems. For the lowest mass systems a large number of spin values are selected. Few systems from the bulk volume of the parameter space are chosen.}
  \label{fig:m1m2spin}
\end{figure}

The 3D equal-spin parameter space ($m_1, m_2, \chi_1=\chi=\chi_2$) can be visualized directly. In Fig.~\ref{fig:m1m2spin} we plot the parameter values selected for the reduced basis. Comparing this to Fig.~\ref{fig:m1m2par} one sees again the usual selection of primarily low-mass systems. For the lowest mass configurations, a large range of spin values is selected. The figure also shows that very little of the bulk of the parameter space volume is used. The greedy algorithm mostly picks values at some edges and corners in the $m_1$-$m_2$ plane and preferentially picks binaries with both spins anti-aligned to the orbital angular momentum ($\chi=-1$); see also Fig.~\ref{fig:s3dhist} showing a histogram of the selected equal-spin magnitudes.

The global aspect of the RB greedy algorithm allows an identification of the underlying sparsity clearly visible in  Fig.~\ref{fig:m1m2spin}. This provides another hint that: 1) the binary coalescence problem is  amenable to dramatic dimensional reduction with respect to the physical parameters describing it, and 2) specific techniques for training space construction exploiting sparsity might provide dramatic computational cost advantages when applied to higher dimensional problems (we elaborate more on this in Sec.~\ref{sec:remarks}).

\subsubsection{Spin parameters (3D, 4D)}

\begin{figure}[htp]
  \centering
  \subfigure[~Spin magnitude distribution (3D case)]{
    \includegraphics[width=0.45\linewidth]{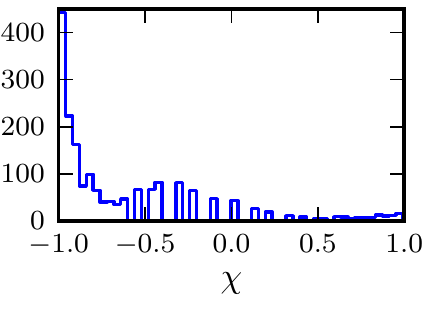}
    \label{fig:s3dhist} }
  \subfigure[~Spin magnitudes distribution (4D case)]{
    \includegraphics[width=0.45\linewidth]{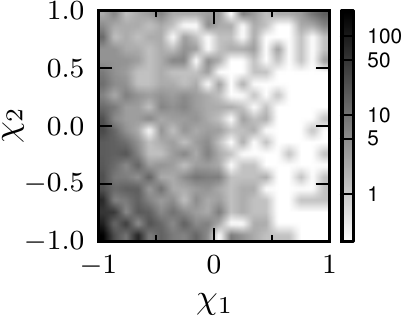}
    \label{fig:s4dhist} }
  \caption{Fig.~\subref{fig:s3dhist} shows a histogram of spin values in the 3D case. Spins anti-aligned with the orbital angular momentum are predominantly chosen. Fig.~\subref{fig:s4dhist} shows a two-dimensional histogram of spin values in the 4D case. Similarly to Fig.~\subref{fig:s3dhist}, anti-aligned systems $\chi_i=-1$ are predominantly chosen. The grey-scale bar utilizes a logarithmic scaling.}
\end{figure}

Fig.~\ref{fig:s4dhist} shows the $1,\!839$ spin magnitudes, $\chi_1$ and $\chi_2$, selected by the greedy algorithm in the generic non-precessing 4D case. The training space corresponding to this figure has $n_s = 20$ points in each spin direction. As found in the 3D case, binaries with both spins mostly anti-aligned with the orbital angular momentum are more relevant than those that are aligned.

The figure is asymmetric in $\chi_1, \chi_2$ for the following reason. Although the representation error of any two waveforms is symmetric in $(m_1, \chi_1) \leftrightarrow (m_2, \chi_2)$, the greedy algorithm needs to select one set of parameters from the two possible choices introducing an asymmetry. This effect is visually enhanced by the logarithmic scale used in the figure.

As discussed in Sec.~\ref{sec:masspars}, the distribution of masses selected does not change strongly as one goes from 2D to 4D. Similarly here, the distribution of spin values selected seems qualitatively similar in the 3D and 4D cases, suggesting that it might be possible to construct training spaces in a more efficient manner.

\subsection{Monte-Carlo representation error studies}\label{sec:reconstruct}

\begin{figure}[htp]
  \includegraphics[width=\linewidth]{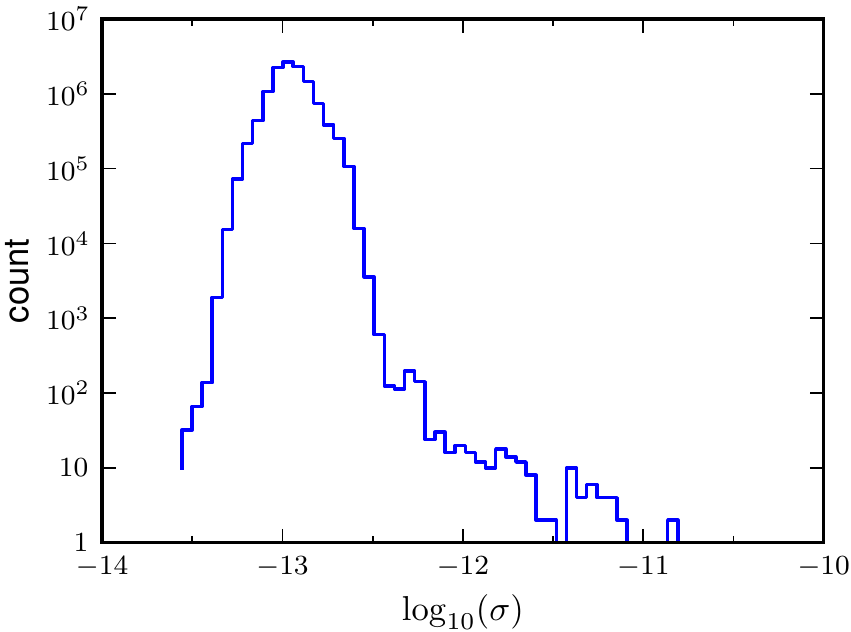}
  \caption{Monte-Carlo Reconstruction error study for the 4D case. The histogram shows the distribution of the representation error $\sigma = \| \tilde{h} - P_{N}[\tilde{h}] \|$ for $10^7$ randomly selected parameter values. Notice the logarithmic scale on the vertical axis.}
  \label{fig:reconstr4d}
\end{figure}

To test the accuracy of the RB to represent waveforms that are not necessarily members of the training space, we randomly sample the corresponding parameter space used to build each basis, and for each sampled waveform we compute its representation error by projecting the waveform onto the RB. We performed this test for all models (2D, 3D, 4D), but for definiteness we show the more interesting 4D case in Fig.~\ref{fig:reconstr4d}. Shown there is the computed waveform representation error for $10^7$ randomly chosen values in the 4D parameter space. The histogram shows that the overwhelming majority of sampled waveforms have a representation error near double-precision round-off values, with a few isolated cases with an error slightly larger than $10^{-11}$. 

For the particular reconstruction test shown in Fig.~\ref{fig:reconstr4d} we used a RB built out of a training space with  $n_m=100$ points for each mass component and $n_s=10$ for each spin magnitude, as described in Sec.~\ref{sec:dimens}. For the 2D case, this number of mass samples is not enough to achieve such a low representation error. This indicates that there is a trade-off between sampling density in the mass and spin dimensions that can be used to reduce the representation error. The effectiveness of the reduced basis approach is such that, for example, even though a relatively small number of spin values are used in the training space to build the RB, a waveform with {\em any} spin value is represented within essentially machine precision. 

\subsection{Seed dependence}

Lastly, we investigate the dependence of the parameters selected by the greedy algorithm on the choice of initial seed for the 2D case (non-spinning). The seeds are chosen to correspond to the  $200$ equal-mass binaries in the training space. For a given seed and tolerance error, the greedy algorithm picks a set of nearly optimal parameters, or a {\it greedy chain}. Different seeds give rise to different greedy chains and the chosen parameters will not necessarily be part of other greedy chains. 

We showed in Ref.~\cite{Caudill:2011kv} that the representation error is robust to different choices of seed values by running the greedy algorithm multiple times, once for each possible value in the training space for the seed~\footnote{Ref.~\cite{Caudill:2011kv} was concerned with quasi-normal mode ringdown waveforms. However, the result is independent of the specific waveform under consideration.}. Here, we discuss how the distribution of points is affected by different seed choices.

\begin{figure}[htp]
  \includegraphics[width=\linewidth]{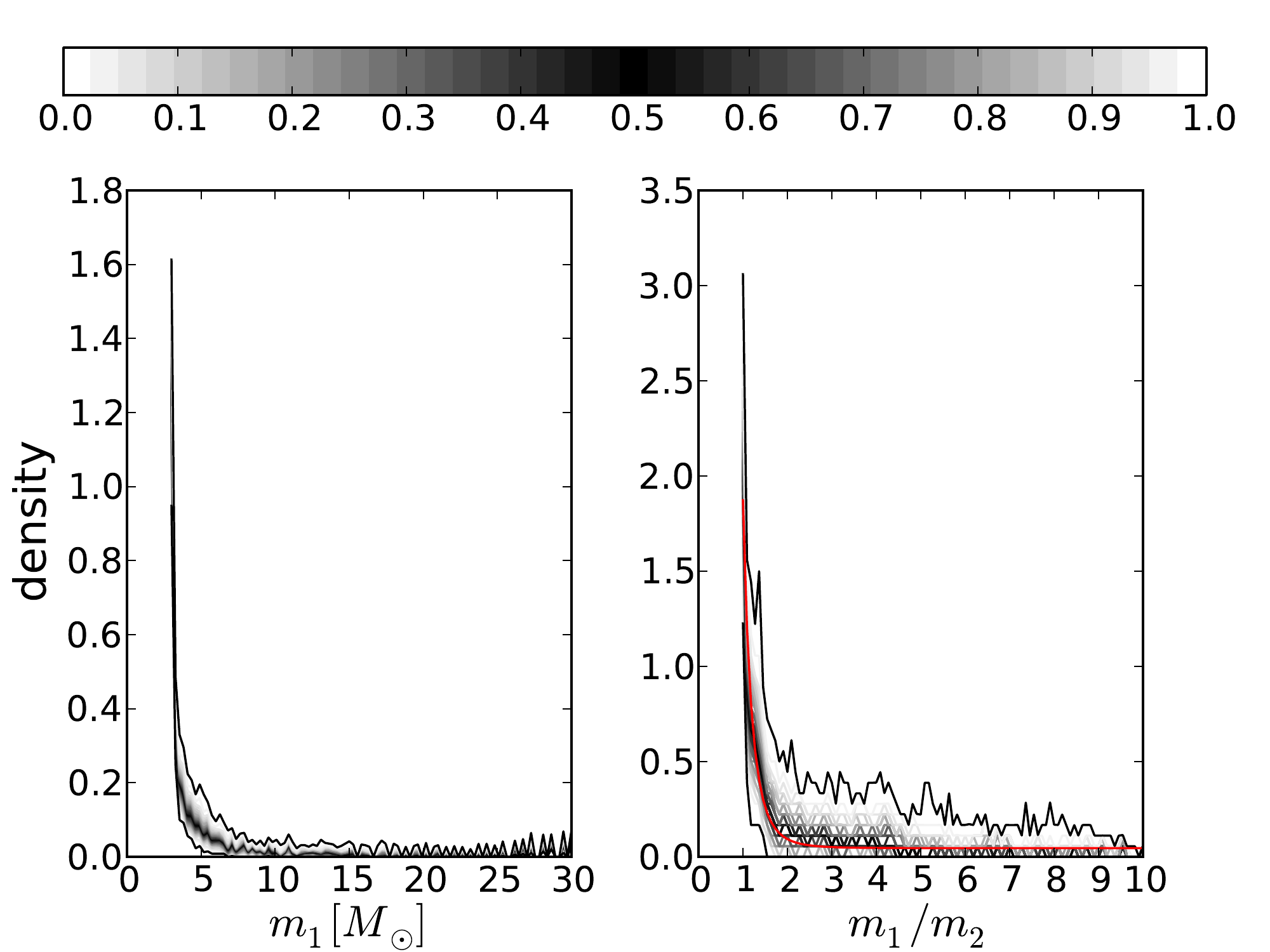}
  \caption{Parameters chosen by the greedy algorithm ($m_1$, left and $m_1/m_2$, right) when using $200$ seed values corresponding to equal mass, non-spinning binaries with different total masses. The lines are quantile curves representing the fraction of selected parameters with values greater than the value of the quantile. Black corresponds to the median and the bounding black curves correspond to the $0\%$ (top) and $100\%$ (bottom) quantiles. The red line is a fit to the median. Despite the different choice of greedy parameters that are selected, the {\it distribution} of points is rather robust to the choice of seed.}
  \label{fig:seedhist}
\end{figure}

Fig.~\ref{fig:seedhist} shows the probability density over all $200$ seeds for a given value of $m_1$ (left) and for $\ratio$ (right), where the mass ratio is taken to be such that if $m_1 > m_2$ then the $1$ and $2$ labels are interchanged so that $\ratio \ge 1$. The gray-level shaded lines correspond to quantile curves, with black indicating the median of the distribution. Both plots indicate that the {\it distribution} of points is rather robust to different seed choices. This reflects the global nature of the greedy algorithm to select parameters that are distributed similarly even though the selected  parameter values themselves exhibit variations from one chain to another. These distributions also show that the majority of points selected, and overwhelmingly so, are those with low masses independently of the seed value. A similar distribution of points exists for the selected values of $m_2$. The red line in the right panel of Fig.~\ref{fig:seedhist} is a power-law fit to the median and is given by $\approx 1.83 \, (m_1/m_2)^{-5.25} + 0.05$. Most mass ratios selected lie within the range $m_1/m_2\in [1,3]$.

\section{Concluding remarks} \label{sec:remarks}

The main focus of this work has been the study of reduced basis in higher dimensional parameter spaces; more specifically, the study of non-precessing, spinning binaries. We have shown (see Figs.~\ref{fig:dimensionality} and \ref{fig:reconstr4d}) that the number of reduced basis vectors needed to represent the full space of non-precessing inspiral waveforms with essentially round-off precision increases very mildly with the number of parameters from 2D to 4D. 

While the work presented here is limited to the non-precessing case, our results provide the first hints, together with \cite{Galley:2010rc} for precessing inspiral dynamics (see also \cite{Pan_etal:PRD69}), \cite{Caudill:2011kv} for quasi-normal mode ringing, and studies using a Singular Value Decomposition (SVD) technique \cite{Cannon:poster,Cannon:2010qh,Cannon:2011xk} that a full representation of the eight-parameter space of inspiral-merger-ringdown waveforms (either numerical, Effective-One-Body, or phenomenological ones)  might actually be achievable with a relatively compact reduced basis. 
If this is the case, then there is hope that the number of simulations needed to represent the space of precessing binary inspirals might be relatively small (perhaps, on the order of several thousand for advanced LIGO but not orders of magnitude larger than this), thus allowing for a tractable number of numerical simulations if the parameters are chosen with malice of forethought from our reduced basis studies. 

Recently, SVD of gravitational waves \cite{Cannon:2010qh} 
has been implemented in a realistic, low-latency search pipeline~\cite{Cannon:2011vi}. 
Such a pipeline could be used to improve upon the work in~\cite{Virgo:2011aa} to allow prompt follow-up searches for electromagnetic counterparts to candidate gravitational wave signals in the advanced detector era. 
As a reconstruction and compression technique, RB has a number of similarities with SVD (e.g., output includes basis vectors and reconstruction coefficients in both schemes). 
Thus the RB results of this paper could be implemented in such a pipeline, in principle. Such practical data analysis implementations of RB will be left for future work.

The exploration of the 8D parameter space will require further development and/or implementation of technical but critical aspects, including the efficient and adaptive sampling techniques for large training spaces (see, e.g., \cite{brown_sc_2011_15}) and rapid evaluation of high accuracy quadratures for parametrized problems \cite{Antil2012}. For the problem at hand, a splitting of dimensions as discussed in Sec.~\ref{sec:masspars} might be useful. 

\section{Acknowledgments} 
This work has been supported by NSF Grants PHY1208861 and PHY1005632 to the University of Maryland, and NSF Grant PHY0970074 to the University of  Wisconsin-Milwaukee. C.\,G. was supported by an appointment to the NASA Postdoctoral Program at the Jet Propulsion Laboratory administered by Oak Ridge Associated Universities through a contract with NASA. Copyright 2012. All rights reserved. 

We thank Alessandra Buonanno, Sarah Caudill, Tom Dent, Alexandre Le Tiec, Drew Keppel, and Dianne O'Leary for very helpful comments on the manuscript and/or suggestions. 

\bibliographystyle{physrev}
\bibliography{references}

\begin{thebibliography}{10}

\bibitem{LIGO_web}
LIGO - http://www.ligo.caltech.edu/.

\bibitem{VIRGO_web}
Virgo - http://www.virgo.infn.it/.

\bibitem{GEO_web}
GEO600 - http://www.geo600.uni-hannover.de/.

\bibitem{KAGRA_web}
KAGRA - http://gwcenter.icrr.u-tokyo.ac.jp/en/.

\bibitem{Abadie:2010cfa}
LIGO Scientific, J.~Abadie {\em et~al.},
\newblock Class. Quantum Grav. {\bf 27}, 173001 (2010), arXiv:1003.2480.

\bibitem{Apostolatos_etal:PRD49}
T.~A. Apostolatos, C.~Cutler, G.~J. Sussman, and K.~S. Thorne,
\newblock Phys. Rev. D {\bf 49}, 6274 (1994).

\bibitem{Brown:2012gs}
D.~A. Brown, A.~Lundgren, and R.~O'Shaughnessy,
\newblock (2012), arXiv:1203.6060.

\bibitem{Ajith:2011ec}
P.~Ajith,
\newblock Phys. Rev. {\bf D84}, 084037 (2011), arXiv:1107.1267.

\bibitem{Centrella:2010zf}
J.~M. Centrella, J.~G. Baker, B.~J. Kelly, and J.~R. van Meter,
\newblock Ann.Rev.Nucl.Part.Sci. {\bf 60}, 75 (2010), arXiv:1010.2165.

\bibitem{Buonanno_etal:PRD70}
A.~Buonanno, Y.~Chen, Y.~Pan, and M.~Vallisneri,
\newblock Phys. Rev. D {\bf 70}, 104003 (2004).

\bibitem{Buonanno:2002fy}
A.~Buonanno, Y.-b. Chen, and M.~Vallisneri,
\newblock Phys. Rev. {\bf D67}, 104025 (2003), arXiv:gr-qc/0211087.

\bibitem{Pan_etal:PRD69}
Y.~Pan, A.~Buonanno, Y.~Chen, and M.~Vallisneri,
\newblock Phys. Rev. D {\bf 69}, 104017 (2004).

\bibitem{Bellman:2003:DP:862270}
R.~E. Bellman,
\newblock {\em Dynamic Programming} (Dover Publications, Incorporated, 2003).

\bibitem{PhysRevD.79.081503}
T.~Damour and A.~Nagar,
\newblock Phys. Rev. D {\bf 79}, 081503 (2009).

\bibitem{Buonanno:1998gg}
A.~Buonanno and T.~Damour,
\newblock Phys. Rev. D {\bf 59}, 084006 (1999), arXiv:gr-qc/9811091.

\bibitem{Pan:2009wj}
Y.~Pan {\em et~al.},
\newblock Phys. Rev. {\bf D81}, 084041 (2010), arXiv:0912.3466.

\bibitem{Taracchini:2012ig}
A.~Taracchini {\em et~al.},
\newblock (2012), arXiv:1202.0790.

\bibitem{Ajith:2007qp}
P.~Ajith {\em et~al.},
\newblock Class. Quantum Grav. {\bf 24}, S689 (2007), arXiv:0704.3764.

\bibitem{Ajith:2009bn}
P.~Ajith {\em et~al.},
\newblock (2009), arXiv:0909.2867.

\bibitem{Santamaria:2010yb}
L.~Santamaria {\em et~al.},
\newblock Phys. Rev. {\bf D82}, 064016 (2010), arXiv:1005.3306.

\bibitem{Field:2011mf}
S.~E. Field {\em et~al.},
\newblock Phys. Rev.Lett. {\bf 106}, 221102 (2011), arXiv:1101.3765.

\bibitem{Caudill:2011kv}
S.~Caudill, S.~E. Field, C.~R. Galley, F.~Herrmann, and M.~Tiglio,
\newblock Class.Quant.Grav. {\bf 29}, 095016 (2012), arXiv:1109.5642.

\bibitem{Cannon:poster}
D.~Keppel, K.~Cannon, M.~Frei, and C.~Hanna,
\newblock http://www.gravity.phys.uwm.edu/conferences/gwpaw/posters/keppel.pdf.

\bibitem{Galley:2010rc}
C.~R. Galley, F.~Herrmann, J.~Silberholz, M.~Tiglio, and G.~Guerberoff,
\newblock Class. Quantum Grav. {\bf 27}, 245007 (2010), arXiv:1005.5560.

\bibitem{Buonanno:2009zt}
A.~Buonanno, B.~Iyer, E.~Ochsner, Y.~Pan, and B.~Sathyaprakash,
\newblock Phys.Rev. {\bf D80}, 084043 (2009), arXiv:0907.0700.

\bibitem{Blanchet:2001ax}
L.~Blanchet, G.~Faye, B.~R. Iyer, and B.~Joguet,
\newblock Phys.Rev. {\bf D65}, 061501 (2002), arXiv:gr-qc/0105099.

\bibitem{Kidder:1995zr}
L.~E. Kidder,
\newblock Phys. Rev. {\bf D52}, 821 (1995), gr-qc/9506022.

\bibitem{Faye:2006gx}
G.~Faye, L.~Blanchet, and A.~Buonanno,
\newblock Phys. Rev. {\bf D74}, 104033 (2006), gr-qc/0605139.

\bibitem{2006PhRvD..74j4034B}
L.~{Blanchet}, A.~{Buonanno}, and G.~{Faye},
\newblock \prd {\bf 74}, 104034 (2006), arXiv:gr-qc/0605140.

\bibitem{Arun:2008kb}
K.~Arun, A.~Buonanno, G.~Faye, and E.~Ochsner,
\newblock Phys. Rev. {\bf D79}, 104023 (2009), arXiv:0810.5336.

\bibitem{AjithFavata}
P.~Ajith and M.~Favata,
\newblock in preparation.

\bibitem{Ajith:2009fz}
P.~Ajith and S.~Bose,
\newblock Phys.Rev. {\bf D79}, 084032 (2009), arXiv:0901.4936.

\bibitem{Binev10convergencerates}
P.~Binev {\em et~al.},
\newblock SIAM J. Math. Analysis {\bf 43}, 1457 (2011).

\bibitem{Cannon:2010qh}
K.~Cannon {\em et~al.},
\newblock Phys. Rev. {\bf D82}, 044025 (2010), arXiv:1005.0012.

\bibitem{Cannon:2011xk}
K.~Cannon, C.~Hanna, and D.~Keppel,
\newblock (2011), arXiv:1101.4939.

\bibitem{Cannon:2011vi}
K.~Cannon {\em et~al.},
\newblock Astrophys.J. {\bf 748}, 136 (2012), arXiv:1107.2665.

\bibitem{Virgo:2011aa}
The LSC and Virgo Collaborations,
\newblock (2011), arXiv:1112.6005.

\bibitem{brown_sc_2011_15}
J.~S. Hesthaven, B.~Stamm, and S.~Zhang,
\newblock Scientific Computing Group, Brown University Report No. 2011-15, 2011
  (unpublished).

\bibitem{Antil2012}
H.~Antil, S.~Field, F.~Herrmann, R.~Nochetto, and M.~Tiglio,
\newblock (2012),
\newblock in preparation.

\bibitem{Chatziioannou:2012rf}
K.~Chatziioannou, N.~Yunes, and N.~Cornish,
\newblock (2012), arXiv:1204.2585.

\bibitem{Yunes:2009ke}
N.~Yunes and F.~Pretorius,
\newblock Phys.Rev. {\bf D80}, 122003 (2009), arXiv:0909.3328.

\end{thebibliography}

\end{document}